\documentclass[twocolumn,prb,showpacs,floatfix,amsfonts,superscriptaddress]{revtex4} 

\usepackage[]{graphicx} 
\usepackage{amsbsy}

\begin{document} 
\title{Ultrafast transient response and electron-phonon coupling in the iron-pnictide superconductor Ba(Fe$_{1-x}$Co$_x$)$_2$As$_2$}

\author{B.~Mansart} 
\affiliation{Laboratoire de Physique des Solides, CNRS-UMR 8502, Universit\'{e} Paris-Sud, F-91405 Orsay, France} 
\author{D.~Boschetto} 
\affiliation{Laboratoire d'Optique Appliqu\'{e}e, ENSTA, CNRS, Ecole Polytechnique, 91761 Palaiseau, France} 
\author{A.~Savoia} 
\affiliation{Laboratoire d'Optique Appliqu\'{e}e, ENSTA, CNRS, Ecole Polytechnique, 91761 Palaiseau, France} 
\author{F.~Rullier-Albenque} 
\affiliation{Service de Physique de l'Etat Condens\'{e}, Orme des Merisiers, CNRS URA 2464, CEA Saclay, 91195 Gif-Sur-Yvette cedex, France} 
\author{F.~Bouquet} 
\affiliation{Laboratoire de Physique des Solides, CNRS-UMR 8502, Universit\'{e} Paris-Sud, F-91405 Orsay, France} 
\author{E.~Papalazarou} 
\affiliation{Laboratoire de Physique des Solides, CNRS-UMR 8502, Universit\'{e} Paris-Sud, F-91405 Orsay, France}
\author{A.~Forget} 
\affiliation{Service de Physique de l'Etat Condens\'{e}, Orme des Merisiers, CNRS URA 2464, CEA Saclay, 91195 Gif-Sur-Yvette cedex, France} 
\author{D.~Colson} 
\affiliation{Service de Physique de l'Etat Condens\'{e}, Orme des Merisiers, CNRS URA 2464, CEA Saclay, 91195 Gif-Sur-Yvette cedex, France} 
\author{A.~Rousse} 
\affiliation{Laboratoire d'Optique Appliqu\'{e}e, ENSTA, CNRS, Ecole Polytechnique, 91761 Palaiseau, France} 
\author{M.~Marsi} 
\affiliation{Laboratoire de Physique des Solides, CNRS-UMR 8502, Universit\'{e} Paris-Sud, F-91405 Orsay, France} 
 
\date{\today} 
 
\begin{abstract} 

The transient response of Ba(Fe$_{1-x}$Co$_x$)$_2$As$_2$, x=0.08 was studied by pump-probe optical reflectivity. After ultrafast photoexcitation, hot electrons were found to relax with two different characteristic times, indicating the presence of two distinct decay channels: a faster one, of less than 1 ps in the considered pump fluence range, and a slower one, corresponding to lattice thermalization and lasting $\cong6 ps$. Our analysis indicates that the fast relaxation should be attributed to preferential scattering of the electrons with only a subset of the lattice vibration modes, with a second moment of the Eliashberg function $\lambda\left\langle\omega^2\right\rangle\cong64~meV^2$. The simultaneous excitation of a strong fully symmetric $A_{1g}$ optical phonon corroborates this conclusion and makes it possible to deduce the value of $\lambda\cong0.12$. This small value for the electron-phonon coupling confirms that a phonon mediated process cannot be the only mechanism leading to the formation of superconducting pairs in this family of pnictides.  

\end{abstract} 
 
\maketitle 
 

The discovery of high temperature superconductivity in iron pnictide compounds in 2008(Refs.~\cite{Kamahira2008} and~\cite{Rotter2008}) has raised a lot of questions about the nature of this phenomenon. One of these questions concerns the role of electron-phonon (e-ph) coupling, which is at the heart of conventional Bardeen-Cooper-Shrieffer (BCS) superconductivity~\cite{Bardeen1957}. 

For unconventional superconductors, electron-lattice interaction mechanisms have been extensively studied in cuprates; these materials present some similarities with pnictides, such as the bidimensionality of the crystallographic structure with Cu-O planes instead of Fe-As ones and the presence of a magnetic phase in the underdoped part of the phase diagram~\cite {McGuire2008, Singh2008}. 
While the electron pairing mechanism is still controversial in superconducting cuprates, some elements are today accepted: first, their high critical temperatures are not compatible with a BCS scheme; second, the e-ph interaction is anisotropic, as theoretically predicted ~\cite{Devereaux2004} and verified by means of time-resolved experiments~\cite{Perfetti2007,Carbone2008, Kusar2008, Giannetti2009, Saichu2009}; indeed, the e-ph coupling constant is strongly mode-selective, and ranges from $\lambda\cong0.13$ to $\lambda\cong0.55$. This selectivity is linked to the marked bidimensional layered structure of cuprates, leading to a preferential coupling between electrons coming from a specific k-direction of the Fermi surface and one particular phonon mode~\cite{Devereaux2004, Carbone2008}. 

Much less information is available on e-ph coupling in pnictides. Theoretical works on the 1111 family (LaFeAsO$_{1-x}$F$_x$), employing Density Functionnal perturbation theory, predicted an isotropic coupling, equally distributed for the whole phonon population, and too small to be responsible for superconductivity through a BCS type mechanism~\cite{Singh2008, Boeri2008}. On the other hand, a very strong coupling ($\lambda\cong1$) between electrons and the A$_{1g}$ mode (consisting of a breathing movement of As atoms) was predicted by a model in which the electronic polarization of As atoms involve e-ph interaction~\cite{Kulic2009}. For 122 compounds (doped $AFe_2As_2$, A=Ba, Sr, Ca) spin fluctuations remaining in the paramagnetic phase may enhance e-ph coupling~\cite{Boeri2010}. As far as experimental results are concerned, the determination of e-ph coupling by pump-probe spectroscopies found a rather low value of $\lambda\cong0.15$ in BaFe$_2$As$_2$~\cite{Chia2010, Stojchevska2010}, raising up to $\lambda\cong0.25$ in SrFe$_2$As$_2$ (Ref.~\cite{Stojchevska2010}) and an intermediate value of $\lambda\cong0.18$ in SmFeAsO~\cite{Mertelj2010}. All of these studies are in defavor of a BCS type coupling as the origin of superconductivity, but noteworthy none of them was performed on doped specimens presenting a superconducting transition.


In this work, we studied the iron pnictide Ba(Fe$_{1-x}$Co$_x$)$_2$As$_2$, at almost optimal doping, x=0.08. This material is superconductor below $T_c$=24 K and doesn't present any magnetic transition~\cite {Rullier-Albenque}. Single crystal samples were grown by the self-flux method~\cite {Rullier-Albenque}, and fully characterized prior to our measurements. 
We carried out pump-probe reflectivity measurements using a mode-locked Ti:Sapphire system delivering laser pulses at 800 nm wavelength and 1 kHz repetition rate. The pulses duration was $\approx 40~fs$. The experimental setup has been described in \cite{Boschetto2008}, allowing us to reach a signal-to-noise ratio up to $10^5$. We kept the linearly polarized pump and probe beams orthogonal to each other; the pump beam was p-polarized, incident at about $5^\circ$ from the surface normal, and was focused on a 100 ${\mu}m$ diameter spot. The probe beam incidence angle was $15^\circ$ with respect to the sample normal, and its size was 25 ${\mu}m$ diameter, in order to probe a uniformly excited area. 
We used pump fluences between $1.3~mJ/cm^{2}$ and $3.2~mJ/cm^{2}$. A continuous flow helium-gas cryostat allowed us to cool the samples down to $\approx 10$ K. The samples were cleaved along the (001) crystallographic direction in order to obtain clean and optically flat surfaces. 


\begin{figure}[!h] 
\includegraphics[angle=0,width=0.8\linewidth,clip=true]{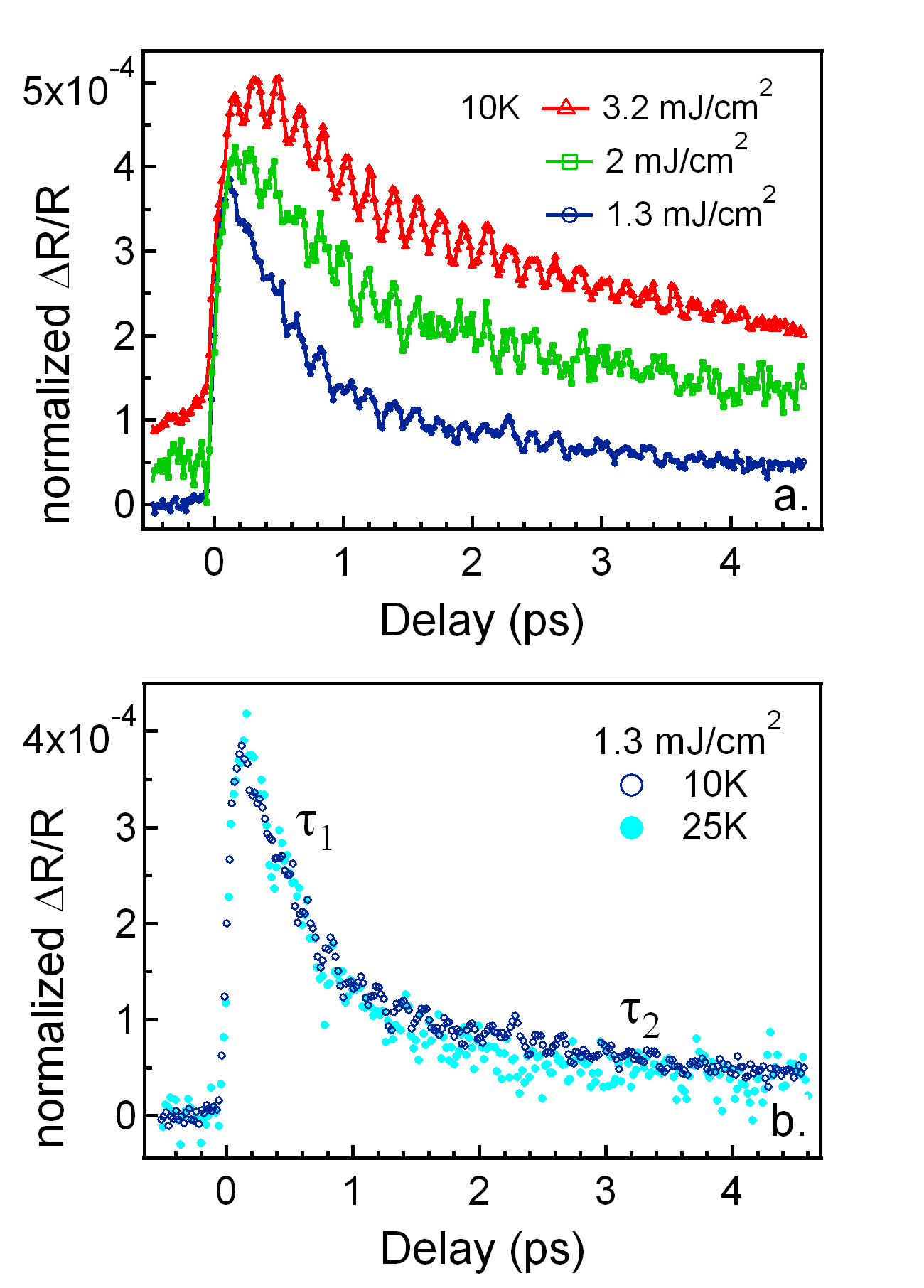} 
\caption{Time-resolved reflectivity curves on Ba(Fe$_{1-x}$Co$_x$)$_2$As$_2$: (a) at $T_i$=10 K (SC phase) for different pump fluences (an offset has been applied for clarity); (b) for $F=1.3~mJ/cm^2$ as a function of $T_i$, across $T_c$. All the curves have been normalized to show the same peak value after the excitation pulse.} 
\label{fig:1} 
\end{figure} 

The time-resolved reflectivity curves are presented in fig. 1, where all the curves have been normalized to have the same value after the pump excitation. The transient reflectivity response consists of a fast increase at zero delay, corresponding to the excitation of electrons by the pump pulse. Then the relaxation dynamics occurs, and the recovery of unperturbed value takes place in several tenths of picoseconds. After the reflectivity increases due to the electron heating, we observe an oscillation, attributed to a coherent A$_{1g}$ mode; its complete study has already been presented in~\cite{Mansart2009}. Here we focus on the transient reflectivity relaxation dynamics occurring on the picosecond time-scale.

This relaxation is composed of two different decay channels; indeed, the curve cannot be described by a single exponential decay, but two of them are necessary. The faster one is hereafter called $\tau_1$ and the slower one $\tau_2$. They are both represented as a function of the maximum electronic temperature $T_e$ (derivated as explained below) in fig. 2, for the two considered initial temperatures $T_i$. $\tau_1$ values range from $\cong500fs$ to $\cong1.1ps$; $\tau_2$ lies between $\cong4.5ps$ and $\cong6.7ps$.
 
\begin{figure}[!h] 
\includegraphics[angle=0,width=0.8\linewidth,clip=true]{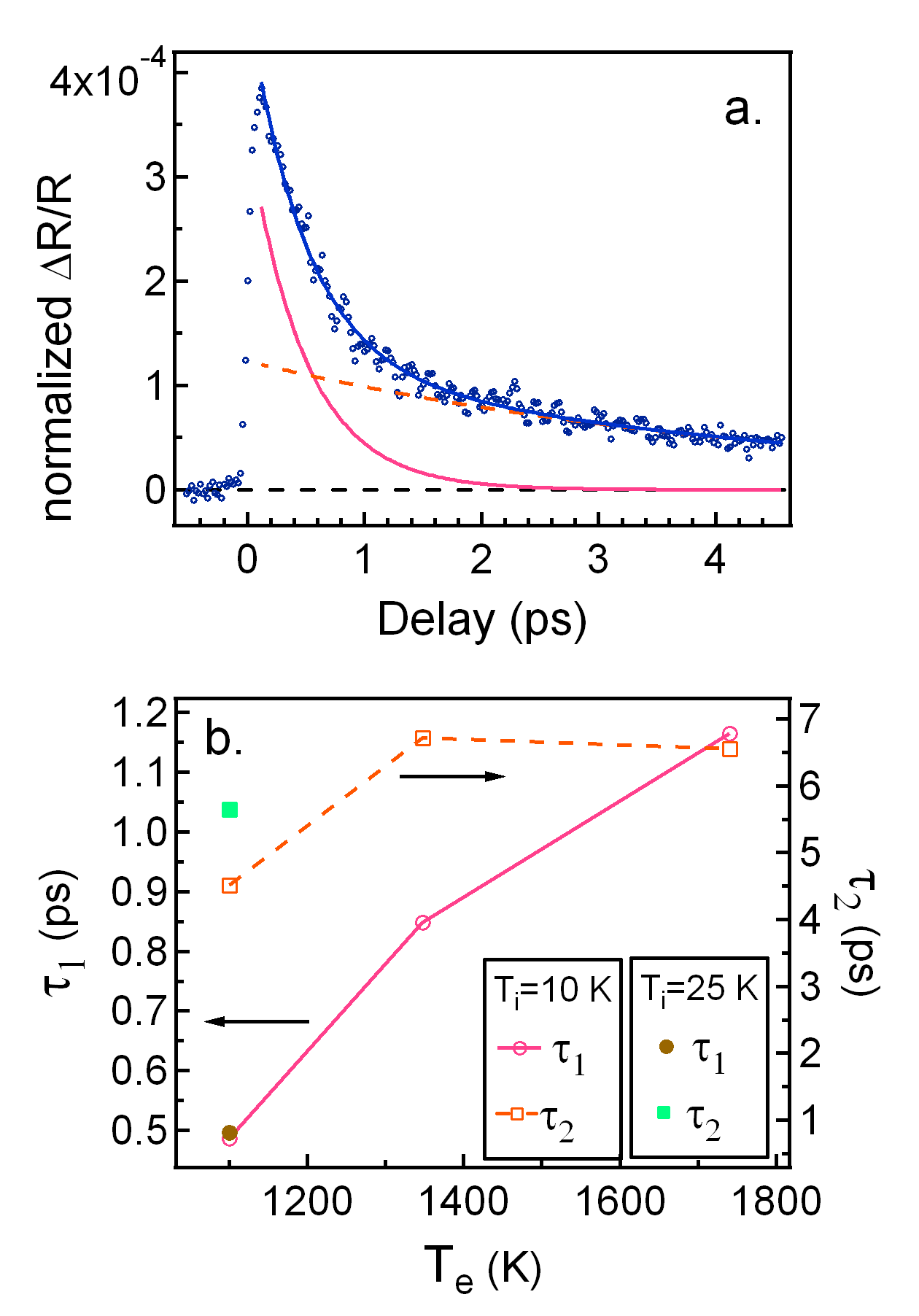} 
\caption{(a) Fitting by a bi-exponential decay function and two components of the fit, for $T_i$=10 K and $F=1.3~mJ/cm^2$; (b) Relaxation times $\tau_1$ and $\tau_2$ as a function of the maximum electronic temperatures.} 
\label{fig:2} 
\end{figure}

Several pump-probe reflectivity measurements performed on iron pnictide family compounds report on the existence of different relaxation times linked to the opening of a gap, either the superconducting (SC) gap or the spin density wave one~\cite{Torchinsky2009, Mertelj2009, Chia2010, Stojchevska2010, Mertelj2010}. This kind of analysis is based on the Rothwarf-Taylor model~\cite{Rothwarf1967}, describing the relaxation of electrons excited from the SC ground state across the gap and their coupling to high energy phonon modes. It should be noted that the pump fluence range used in this work is much higher (by several orders of magnitude) than all the previous studies. In this excitation energy density range (~$mJ/cm^2$) the measurements are sensitive to the relaxation due to electron-phonon coupling, while for lower fluences (~$\mu J/cm^2$) one has direct access to Cooper pair recombination.  

The strong excitation level effect can be seen by calculating the electronic temperature reached just after excitation. This can be done as a function of the pump fluence $F$ by (see for example~\cite{Boschetto2008} and references therein):
 \begin{equation}
\label{Te vs F}
 T_e=\left\langle\sqrt{T_i^2+\frac{2(1-R)F}{l_s\gamma}e^{-z/l_s}}\right\rangle
\end{equation}
where R$\cong0.32$ at 1.55 eV~\cite{Drechsler2009} is the unperturbed reflectivity and $\gamma$ the linear coefficient of heat capacity due to the electronic sub-system. The mean value is taken for the depth z going from the sample surface down to the skin depth $l_s$=26nm (we used the value measured in Ba$_{0.6}$K$_{0.4}$Fe$_2$As$_2$~\cite{Li2008}). Prior to our time-resolved reflectivity measurements, we performed high resolution, low temperature heat capacity measurements in samples coming from the same batches. The results are in excellent agreement with previous studies~\cite{Hardy2010, Storey2010}. For $T_c<T<<\Theta_D$ ($\Theta_D\cong300 K$~\cite{Storey2010} is the Debye temperature), the heat capacity behaves as $C(T)=\gamma T+\beta T^3$. We then obtained the accurate electronic heat capacity $C_e=\gamma T_e$, with $\gamma=21~mJ/mol/K^2$ and the lattice one $C_V=\beta T_l^3$, with $\beta=0.40~mJ/mol/K^4$.

The maximum $T_e$ ranges from 1100 K to 1740 K, far above the initial temperature. In Ba(Fe$_{1-x}$Co$_x$)$_2$As$_2$ and the pump fluence range considered here, the second term under the square root of (1) is dominant with respect to the first one, $T_i$; therefore, $T_e$ is much more strongly affected by variations of $F$ than by these of $T_i$.

The fact that $\tau_1$ is unchanged across the superconducting phase transition (fig. 1 (b)) is a direct proof that it is not linked to a electronic relaxation across a gap, as expected considering the high $T_e$. Moreover, its rather linear dependence as a function of $T_e$ (fig. 2) reflects the behavior of an e-ph relaxation time~\cite{Allen1987}. On the other hand, the slow relaxation time $\tau_2$ slightly increases as a function of $T_e$. 

The presence of two decay times suggests a preferential coupling between electrons and some phonon modes. As in cuprates, the marked layered crystallographic structure could induce such effects, and it is actually theoretically predicted to occur with the $A_{1g}$ mode coherently excited here~\cite{Kulic2009}, where As atoms move perpendicularly to the Fe-As plane. Excited electrons first thermalize over the time scale $\tau_1$ with the more coupled phonon modes, which in turn relax their excess energy over the time-scale $\tau_2$. The most likely mechanism for this second decay
time is thermalization with the remaining phonon modes; spin-lattice interactions 
appear unlikely, since the x=0.08 doped sample under study doesn't show any magnetic transition, but can't be completely ruled out due to the presents antiferromagnetic fluctuations.

 In order to extract the e-ph coupling constant, we used a three temperature model to simulate the evolution of electrons, 'preferential phonons' and 'remaining phonons' as a function of the delay time. 

The three temperature model equations are~\cite{Allen1987, Perfetti2007}:
\begin{equation}
\begin{array}{l}
\label{Te vs t}
 2 C_e \frac{\partial T_e}{\partial t}=\frac{2(1-R)}{l_s}I(t)-g(T_e-T_1)\\
 ~\\
 C_1 \frac{\partial T_1}{\partial t}=g(T_e-T_1)-g_l(T_1-T_2)\\
 ~\\
 C_2 \frac{\partial T_2}{\partial t}=g_l(T_1-T_2)\\
 \end{array}
\end{equation}

where indexes $e$ stand for electrons, $1$ for 'prefentially coupled phonons' and $2$ for the rest of phonons. I(t) is the laser intensity, exciting electrons only at time zero; $g$ is the e-ph coupling constant which governs the fast decaying exponential, and $g_l$ the phonon-phonon diffusion governing the slow one. Phonon heat capacities $C_1$ and $C_2$ are taken to be partial of the total lattice one, i.e. $C_1=\alpha C_V$ and $C_2=(1-\alpha) C_V$, $\alpha$ representing the fraction of preferentially coupled phonons. The whole temperature range dependence of $C_V$ has been taken into account by performing interpolation between the measured low-temperature behavior and the high temperature values following the experimental curve in~\cite{Storey2010}. We neglected electron diffusion because of the quasi-bidimensionality of the system, which reduces thermal diffusion along the (001) direction.
Once the evolutions of $T_e$, $T_1$ and $T_2$ are known, we estimate the change of reflectivity to be a linear combination of electronic and average lattice temperature $\left\langle T_{latt}\right\rangle=\alpha T_1 + (1-\alpha) T_2$, following the theoretical derivation in Ref.~\cite{Boschetto2008},
\begin{equation}
\begin{array}{l}
\label{damped harmonic oscillator}
\frac{\Delta R (t)}{R} = A_{e}T_{e}(t)+A_{latt}\left\langle T_{latt}(t)\right\rangle
 \end{array}
\end{equation}

\begin{figure}[t!] 
\includegraphics[angle=0,width=1\linewidth,clip=true]{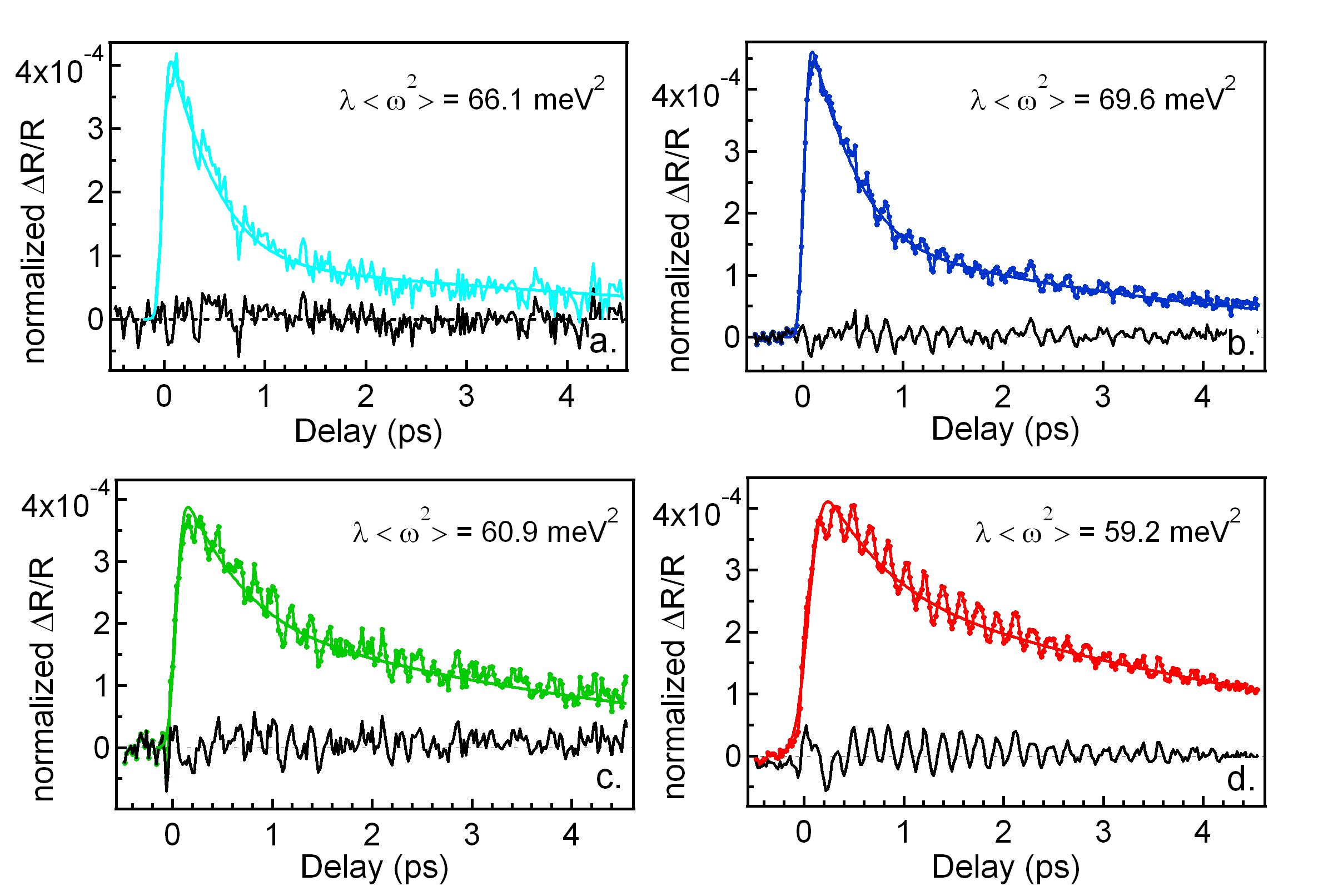} 
\caption{Three temperatures model simulations of the transient reflectivity curves. (a) $T_i$=25 K, $F=1.3~mJ/cm^2$ (b) $T_i$=10 K, $F=1.3~mJ/cm^2$ (c) $T_i$=10 K, $F=2~mJ/cm^2$ (d) $T_i$=10 K, $F=3.2~mJ/cm^2$. The difference between experimental curve and simulation are shown in black.} 
\label{fig:3} 
\end{figure} 

\begin{table}[!h]
\begin{tabular}{|c|c|c|c|c|c|}
\hline
$T_i$ & $F$ & $T_e^{max}$ & $\alpha$ & $\lambda \left\langle \omega^2 \right\rangle$ & $\lambda$\\
 (K) & $(mJ/cm^2)$ & (K) & & ($meV^2$)&\\
\hline
25  & 1.27 & 1100 & 0.4& 66.1 & 0.125\\
\hline
10 & 1.27 & 1100 & 0.4& 69.6 & 0.132\\
   & 1.91 & 1350 & 0.35& 60.9 & 0.115\\
   & 3.18 & 1740 & 0.35& 59.2 & 0.112\\
\hline
		\end{tabular}
	\caption{Maximum electronic temperatures, fraction of strongly coupled phonons and electron-phonon coupling constants obtained by Three Temperature Model simulations.}
\end{table}

Finally, the simulations are convolved with a gaussian to take into account the finite pulse duration. Results of simulation and comparison with experimental data are given in fig. 3, where we can notice a good agreement between experiments and theory. The relevant parameters used to simulate the transient reflectivity are given in table 1. 
The slight disagreement at low delay time between simulation and experimental curve at the highest pump fluence (fig. 3 (d)) comes probably from the electronic heat diffusion. This parameter has not been taken into account in our equations, and would increase with the electronic temperature, enlarging the curve just after excitation. 

The obtained fraction of preferentially coupled phonons lies between 0.35 and 0.4, which is higher than the value 0.2 reported in cuprates~\cite{Perfetti2007}. An explanation for this may be related to the bidimensional structure of these materials, which may be an important factor in the selection of some phonon modes to be more efficiently coupled with electrons. 
The crystallographic structure of iron pnictides is much less bidimensionnal than Bi$_2$Sr$_2$CaCu$_2$O$_{8+\delta}$, especially for Co-doped BaFe$_2$As$_2$ (see for example ~\cite{Vilmercati2009, Brouet2009}): therefore the selectivity of phonons is less marked, and a more homogeneous coupling is observed. We notice here that the presence of two distinct decay times in SrFe$_2$As$_2$ (Ref.~\cite{Stojchevska2010}) also lead to the conclusion of selective e-ph coupling. 

As a result of our analysis, we obtain an e-ph coupling $6.8<g<$$8.0~mJ/K/s/m^3$. By using the relation $g=\frac{6\hbar \gamma}{\pi k_B} \lambda \left\langle \omega^2 \right\rangle$~\cite{Allen1987}, we obtain for the second moment of the Eliashberg function $\lambda \left\langle \omega^2 \right\rangle \cong 64~meV^2$, in good agreement with Refs.~\cite{Chia2010} and~\cite{Stojchevska2010}. 

In the estimation of $\left\langle \omega^2\right\rangle$, we also have to take into account the fact that some vibrational modes are more efficiently coupled than others. The most natural choice is to take the frequency of the fully symmetric $A_{1g}$ mode, which is coherently excited by our photoexcitation 
and consequently efficiently coupled. Moreover, it is the exact counterpart of the preferentially coupled mode of Bi$_2$Sr$_2$CaCu$_2$O$_{8+\delta}$ (the so-called buckling mode, where O atoms oscillate othogonally to the Cu-O layer~\cite{Devereaux2004, Carbone2008}). Then we used the $A_{1g}$ energy, 23 meV~\cite{Mansart2009}, as $\left\langle \hbar \omega\right\rangle$, which gives $\lambda \cong 0.12$. We notice that, by taking the average of the whole phonon spectrum as measured by neutron diffusion~\cite{Mittal2008}, we obtain an even smaller but very close value of $\lambda \cong 0.10$; the coupling given here is therefore a higher limit estimation of $\lambda$.

With this value for $\lambda$, we can evaluate the critical temperature $T_c$ using the McMillan formula~\cite{McMillan1968}, valid for moderate e-ph coupling and isotropic system:

\begin{equation}
\label{Tc vs lambda}
T_c =\frac{\left\langle \omega \right\rangle}{1.20}exp\left[-\frac{1.04(1+\lambda)}{\lambda-\mu^*(1+0.62\lambda)}\right]
\end{equation}

By taking $\left\langle \hbar \omega\right\rangle$=23 meV and $\mu^*$=0~\cite{ref33}, we obtain $T_c\cong0.03K$, which is far below the actual $T_c$ of about 24 K: therefore, the electron pair formation cannot be explained only by electron-lattice interactions. Our results confirm that the scenario of an e-ph mediated superconductivity mechanism can be ruled out for these compounds ~\cite{Boeri2008}.


In conclusion, we reported an ultrafast transient reflectivity study on optimally doped Ba(Fe$_{1-x}$Co$_x$)$_2$As$_2$. The fact that two different time scales need to be taken into account to describe the relaxion dynamics, as well as the simultaneous observation of a strong coherently excited optical phonon, indicate that a preferential coupling exists with some vibrational modes. On the basis of our analysis, we can estimate a value of $\lambda \left\langle \omega^2 \right\rangle\cong64~meV^2$: if we consider a more efficient coupling with low energy modes, and especially with the coherently excited $A_{1g}$ one, a value for the e-ph coupling constant of $\cong0.12$ is obtained. This low value cannot explain a $T_c$ of 24 K within a standard BCS framework, and confirms that other kind of mechanisms are responsible for superconductivity in these materials.


The authors gratefully acknowledge G. Rey for his expert help with the laser system, L. Perfetti for stimulating discussions, and P. A. Albouy for generous loan of equipment.

\end{document}